\documentclass[pre,onecolumn,10pt]{revtex4}%
\usepackage{amsfonts}
\usepackage{amsmath}
\usepackage{amssymb}
\usepackage{graphicx}%
\begin{document}
\title{The 3-dimensional cube is the only periodic, \\connected cubic graph with perfect state transfer}
\author{Simone Severini}
\email{simoseve@gmail.com}
\affiliation{Department of Physics and Astronomy, University College London, WC1E 6BT
London, United Kingdom}

\begin{abstract}
There is \emph{perfect state transfer} between two vertices of a graph, if a
single excitation can travel with fidelity one between the corresponding sites
of a spin system modeled by the graph. When the excitation is back at the
initial site, for all sites at the same time, the graph is said to be
\emph{periodic}. A graph is \emph{cubic} if each of its vertices has a
neighbourhood of size exactly three. We prove that the 3-dimensional cube is
the only periodic, connected cubic graph with perfect state transfer. We
conjecture that this is also the only connected cubic graph with perfect state transfer.

\end{abstract}
\maketitle

\section{Introduction}

\subsection{State transfer}

Let $G=(V,E)$ be a graph with set of vertices $V(G)=\{1,2,...,n\}$ and set of
edges $E(G)\subseteq V(G)\times V(G)-\{\{i,i\}:i\in V(G)\}$. The \emph{order}
of $G$ is the number of its vertices.

Let us consider a system of $n$ spin-$1/2$ quantum particles with unit $XY$
couplings. The space assigned to the entire system is $\left(  \mathbb{C}%
^{2}\right)  ^{\otimes n}$. Each particle is attached to a vertex of $G$. The
coupling between two particles is nonzero if and only if the particles are
attached to adjacent vertices. The couplings are then specified by the
adjacency matrix of the graph. The $ij$-th entry of the \emph{adjacency
matrix} of $G$ is $[A(G)]_{i,j}=1$ if $\{i,j\}\in E(G)$ and $[A(G)]_{ij}=0$ if
$\{i,j\}\notin E(G)$.

We shall work with the XY model. Let $X_{i}$ and $Y_{i}$ be the Pauli
operators acting on the $i$-th particle. The Hamiltonian governing the
dynamics of the spin system can be written as $H_{XY}(G)=2^{-1}\sum_{i\neq
j=1}^{n}[A(G)]_{i,j}\left(  X_{i}X_{j}+Y_{i}Y_{j}\right)  $. Let
$\{|1\rangle\equiv\mathbf{e}_{1},|2\rangle,\ldots,|n\rangle\}$ be the standard
basis of the space $\mathbb{C}^{n}$. A vector $|i\rangle$ indicates the
presence of an excitation at vertex $i$ only. With respect to the standard
basis, the $ij$-th entry of the Hamiltonian acting on $\mathbb{C}^{n}$ is
$[H_{XY}(G)]_{i,j}=2[A(G)]_{i,j}$. It follows that the Schr\"{o}dinger
evolution of the excitation is practically induced by a unitary matrix of the
form $U_{G}(t)=e^{-iA(G)t}$, where $t\in\mathbb{R}^{+}$. We obtain a
probability distribution supported by $V(G)$ by performing a projective
measurement on the state $|\psi_{t}\rangle=U_{G}(t)|\psi_{0}\rangle$. For
regular graphs, the \textquotedblleft practically\textquotedblright\ has a
much larger extension, given that, with constant couplings, any kind of
interaction has an Hamiltonian proportional to $A(G)$.

Given two vertices $i,j\in V(G)$, the \emph{fidelity} at time $t$ between $i$
and $j$ is the function $f_{G}(i,j;t)=\left\vert \langle j|U_{G}%
(t)|i\rangle\right\vert $. We say that there is \emph{perfect state transfer}
(for short, \emph{PST}) between the particles $i$ and $j$ at time $t$ if
$f_{G}(i,j;t)=1$ \cite{ch}. We say that $G$ is \emph{periodic}, with period $t
$, if $f_{G}(i,i;t)=1$ \cite{god}. Sometime in the physics literature periodic
graphs are said to afford \emph{perfect revival} (see, \emph{e.g.}, \cite{blu}).

Even if the topic is not directly discussed here, it is worth noticing that in
the $XYZ$ model, the Hamiltonian restricted to $\mathbb{C}^{n}$ is
proportional to the Laplacian matrix of $G$ (see, \emph{e.g.}, \cite{bos1}).
This fact alone is sufficient to distinguish different approaches for the two
models, when the graphs considered have generic degree sequences. In this
work, we consider regular graphs only. The result obtained is then also valid
for the $XYZ$ model.

The concept of PST has been introduced in \cite{bo} and \cite{ch}. The recent
papers \cite{god1} and \cite{an}, even if not reviews, point out a good number
of references embracing the more mathematical aspects around the notion.

\subsection{Diameter}

A graph $H=(W,F)$ is a \emph{subgraph} of $G$ if $W(H)\subseteq V(G)$ and
$F(H)\subseteq E(G)$. A subgraph $H=(W,F)$ is an \emph{induced}
\emph{subgraph} of $G$ if $H$ is a subgraph of $G$ and, for every two vertices
$i,j\in V(H) $, $\{i,j\}\in E(H)$ if and only if $\{i,j\}\in E(G)$.

The \emph{degree} of a vertex $i$ is the number of edges incident with $i$. A
\emph{path} of length $l$ \emph{from} vertex $i$ \emph{to} vertex $j$ (if
there is one) is an induced subgraph with $l$ vertices and $l-1$ edges, such
that $i$ and $j$ have degree one and all other vertices in the path have
degree two. A graph is said to be \emph{connected} if every two vertices are
in a path.

Let $\mathcal{P}_{i,j}(G)$ be the set of all paths with end-vertices $i$ and
$j$. The \emph{length} of a path with end-vertices $i$ and $j$ is denoted by
$l(i,j)$. The (geodesic) \emph{distance} between two vertices $i$ and $j$ is
defined as $d(i,j)=\min_{\mathcal{P}_{i,j}(G)}l(i,j)$. The \emph{diameter} of
a connected graph $G$ is defined as dia$(G)=\max_{i,j\in V(G)}d(i,j)$.
Informally, the diameter is the longest of the shortest paths. Two vertices
$i,j\in V(G)$ are said to be \emph{antipodal} if $d(i,j)=$ dia$\left(
G\right)  $.

\subsection{Order/distance problems for state transfer}

There is a large and growing literature concerned with the mathematics of
state transfer on spin systems. Some effort towards a classification may be
seen as driven by three recurrent, but essentially unstated problems:

\begin{itemize}
\item \emph{General graphs:}\textbf{\ }Given $D\in\mathbb{N}$, find the graph
$G=(V,E)$ with the smallest possible number of vertices $n_{D}=\left\vert
V(G)\right\vert $ such that $d(i,j)=D$ and $f_{G}(i,j;t)=1$ for some
$t\in\mathbb{R}^{+}$. The set of these graphs is denoted by $\mathcal{G}_{D}$.

\item \emph{Fixed degree graphs:}\textbf{\ }Given $D,\Delta\in\mathbb{N}$,
find the graph $G=(V,E)$ with the smallest possible number of vertices
$n_{D,\Delta}=\left\vert V(G)\right\vert $ such that \emph{(i) }the maximum
degree of $G$ is $\Delta$, \emph{(ii) }$d(i,j)=D$ and $f_{G}(i,j;t)=1$ for
some $t\in\mathbb{R}^{+}$. The set of these graphs is denoted by
$\mathcal{G}_{D,\Delta}$.

\item \emph{Regular graphs:}\textbf{\ }Given $D,k\in\mathbb{N}$, find the
graph $G=(V,E)$ with the smallest possible number of vertices $n_{D,k}%
=\left\vert V(G)\right\vert $ such that \emph{(i) }$G$ is $k$-regular,
\emph{(ii) }$d(i,j)=D$ and $f_{G}(i,j;t)=1$ for some $t\in\mathbb{R}^{+}$. The
set of these graphs is denoted by $\mathcal{G}_{D,k}^{R}$. A graph is
$k$-\emph{regular }if all of its vertices have degree $k$.
\end{itemize}

The requirement \textquotedblleft smallest possible number\textquotedblright%
\ could be replaced with \textquotedblleft largest possible
number\textquotedblright\ to state the specular versions of the problems.

\subsection{Examples}

\begin{itemize}
\item $D=1$: Let $P_{2}=$ (\{1,2\},\{\{1,2\}\}) be the path of length one.
Then $[U_{P_{2}}\left(  t\right)  ]_{1,2}=-i\sin\left(  t\right)  $ and
\[
\max_{t\in\mathbb{R}^{+}}(\left\vert [U_{P_{2}}\left(  t\right)
]_{1,2}\right\vert )=f_{P_{2}}\left(  1,2;\pi/2\right)  =1.
\]
Hence, $P_{2}\in\mathcal{G}_{1}$.

\item $D=2$: Let $P_{3}=$ (\{1,2,3\}, \{\{1,2\}, \{2,3\}\}) be the path of
length two. Then $[U_{P_{3}}\left(  t\right)  ]_{1,3}=-\sin\left(  t/\sqrt
{2}\right)  ^{2}$ and
\[
\max_{t\in\mathbb{R}^{+}}\left(  \left\vert [U_{P_{3}}\left(  t\right)
]_{1,3}\right\vert \right)  =f_{P_{3}}\left(  1,3;\pi/\sqrt{2}\right)  =1.
\]
Hence, $P_{3}\in\mathcal{G}_{2}$.
\end{itemize}

In both cases, PST is between antipodal vertices.

\begin{itemize}
\item $D=3$: Let $P_{4}=$ (\{1,2,3,4\}, \{\{1,2\}, \{2,3\}, \{3,4\}\}) be the
path of length three. Let $a:=t/2$. Then $[U_{P_{4}}\left(  t\right)
]_{1,4}=i\sqrt{5}\sin\left(  a+a\sqrt{5}\right)  /10-i\sin\left(  a+a\sqrt
{5}\right)  /2-i\sqrt{5}\sin\left(  a-a\sqrt{5}\right)  /10-i\sin\left(
a-a\sqrt{5}\right)  /2$ and
\[
\max_{t\in\mathbb{R}^{+}}\left(  \left\vert [U_{P_{4}}\left(  t\right)
]_{1,4}\right\vert \right)  =f_{P_{4}}\left(  1,3;2\pi/\sqrt{5}\right)
=\sin\left(  \pi/\sqrt{5}\right)  \approx0.986.
\]
Hence, $P_{4}\notin\mathcal{G}_{3}$.
\end{itemize}

\subsection{Cartesian products}

The \emph{Cartesian product} $G=G_{1}\times G_{2}=(V,E)$ of two graphs
$G_{1}=\left(  V_{1},E_{1}\right)  $ and $G_{2}=\left(  V_{2},E_{2}\right)  $
has set of vertices $V(G)=V(G_{1})\times V(G_{2})$ and $\{\{i,j\},\{k,l\}\}\in
E\left(  G\right)  $ if \emph{(i) }$i=k$ and $\{j,l\}\in E(G_{2})$ or
\emph{(ii)} $j=l$ and $\{i,k\}\in E(G_{1})$. Two facts are important:
dia$\left(  G\right)  =$ dia$\left(  G_{1}\right)  +$ dia$\left(
G_{2}\right)  $; if $G_{1}$ and $G_{2}$ are $k$-regular and $l$-regular
graphs, respectively, then $G$ is $(k+l)$-regular.

\begin{figure}
[ptb]
\begin{center}
\includegraphics[
height=1.1344in,
width=3.6322in
]%
{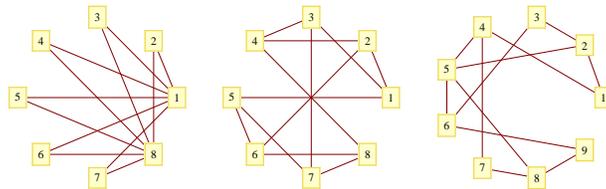}%
\caption{\emph{(L)}: The graph $W$. There is PST between vertices $1$ and $8$.
The spectrum of $W$ is not integral. \emph{(C)}: The 3-dimensional cube. There
is PST between vertices $1$ and $8$. These vertices are antipodal and at
distance $3$. \emph{(R)}: The graph $P_{3}^{\times2}$. Note that the graph has
vertices of degree two and three. There is PST between the vertices $1$ and
$9$. These vertices are antipodal and at distance four. }%
\label{graphprima}%
\end{center}
\end{figure}

The $k$\emph{-dimensional cube} is the graph $P_{2}^{\times k}$ (in Fig.
\ref{graphprima}, $P_{2}^{\times3}$). We have the following four cases:%
\[
\lbrack U_{P_{2}^{\times k}}\left(  t\right)  ]_{1,2^{k}}=\left\{
\begin{tabular}
[c]{rr}%
$-\sin\left(  t\right)  ^{k},$ & if $k=2l$, $l$ odd;\\
$\sin\left(  t\right)  ^{k},$ & if $k=2l$, $l$ even;\\
$i\sin\left(  t\right)  ^{k},$ & if $k=2l+1$, $l$ odd;\\
$-i\sin\left(  t\right)  ^{k},$ & if $k=2l+1$, $l$ even.
\end{tabular}
\ \right.
\]
In all these cases,
\[
\max_{t\in\mathbb{R}^{+}}\left(  \left\vert [U_{P_{2}^{\times k}}\left(
\pi/2\right)  ]_{1,2^{k}}\right\vert \right)  =f_{P_{2}^{\times k}}\left(
1,2^{k};\pi/2\right)  =1.
\]
For $P_{2}^{\times k}$, we have $\left\vert V(P_{2}^{\times k})\right\vert
=2^{k}$ and dia$(P_{2}^{\times k})=k\cdot$ dia$(P_{2})=2\log_{2}\left\vert
V(P_{2}^{\times k})\right\vert =k$. Thus, for the $k$-regular graphs in
$\mathcal{G}_{m,m}^{R}$, $n_{m,m}\leq2^{m}$, where $m=D,k$.

The $k$-dimensional generalization of the $3\times3$ grid is denoted by
$P_{3}^{\times k}$ (in Fig. \ref{graphprima}, the grid $P_{3}^{\times3}$).
There are four types of matrix entries for the graph $P_{3}^{\times k}$:%
\[
\lbrack U_{P_{3}^{\times k}}\left(  t\right)  ]_{1,3^{k}}=\left\{
\begin{tabular}
[c]{rr}%
$-\sin\left(  t/\sqrt{2}\right)  ^{2k},$ & if $k=2l$, $l$ odd;\\
$\sin\left(  t/\sqrt{2}\right)  ^{2k},$ & if $k=2l$, $l$ even;\\
$i\sin\left(  t/\sqrt{2}\right)  ^{2k},$ & if $k=2l+1$, $l$ odd;\\
$-i\sin\left(  t/\sqrt{2}\right)  ^{2k},$ & if $k=2l+1$, $l$ even.
\end{tabular}
\right.
\]
Then,
\[
\max_{t\in\mathbb{R}^{+}}\left(  \left\vert [U_{P_{3}^{\times k}}\left(
\pi/\sqrt{2}\right)  ]_{1,3^{k}}\right\vert \right)  =f_{P_{3}^{\times3}%
}\left(  1,3^{k};\pi/\sqrt{2}\right)  =1.
\]
For $P_{3}^{\times k}$, we have $\left\vert V(P_{3}^{\times k})\right\vert
=3^{k}$ and dia$(P_{3}^{\times k})=k\cdot$ dia$(P_{3})=2\log_{3}\left\vert
V(P_{3}^{\times k})\right\vert =2k$.

For $P_{2}^{\times k}$ and $P_{3}^{\times k}$ PST is between antipodal
vertices. The parameters related to PST between two vertices $i$ and $j$ in
these graphs are given in the following tables \cite{ch}:%
\[
\begin{tabular}
[c]{l|l|l|l}%
$P_{2}^{\times k}$ & $k$ & $\left\vert V(P_{2}^{\times k})\right\vert $ &
$d(i,j)$\\\hline
& $2$ & $4$ & $2$\\
& $3$ & $8$ & $3$\\
& $4$ & $16$ & $4$\\
& $5$ & $32$ & $5$%
\end{tabular}
\text{;\ }%
\begin{tabular}
[c]{l|l|l|l}%
$P_{3}^{\times k}$ & $k$ & $\left\vert V(P_{3}^{\times k})\right\vert $ &
$d(i,j)$\\\hline
& $2$ & $9$ & $4$\\
& $3$ & $27$ & $6$\\
& $4$ & $81$ & $8$\\
& $5$ & $243$ & $10$%
\end{tabular}
\text{ }.
\]
Notice that $P_{3}^{\times k}$ is nonregular for every $k$.

A square matrix $M$ of size $n$ consisting of unimodular entries $\left\vert
M_{i,j}\right\vert =1$ is called a \emph{Hadamard matrix} if $HH^{\dagger}=nI
$, where $I$ is the identity matrix and $^{\dagger}$ denotes the Hermitian
transpose. In a \emph{complex Hadamard matrix}, $M_{i,j}\in\mathbb{C}$
\cite{ka}. The matrix
\[
U_{P_{2}^{\times k}}\left(  \frac{\pi}{4}\right)  =\frac{1}{\sqrt{2}}\left[
\begin{array}
[c]{rr}%
1 & -i\\
-i & 1
\end{array}
\right]  ^{\otimes k}
\]
is a complex Hadamard matrix.

\subsection{Statement of the results}

We prove that the 3-dimensional cube is the only periodic, connected cubic
graph with PST. Equivalently,

\bigskip

\noindent\textbf{Theorem. }The $3$-dimensional cube, $P_{2}^{\times3}$, is the
only periodic, connected cubic ($3$-regular) graph $G$ with two different
vertices $i$ and $j$ such that $f_{G}(i,j;t)=1$, for some $t\in\mathbb{R}^{+}$.

\bigskip

The statement is verified directly in the next section. Conclusions follow.
The proof is based on two known results: a periodic, connected regular graph
is integral; there are only thirteen connected cubic integral graphs. The
proof is technically easy, but tedious, because it goes through a seemingly
unavoidable case by case analysis. Nonetheless, establishing the result is
also an excuse for a further step into a systematical exploration of periodic
quantum dynamics. The proof is interspersed with extra information. The
broader aim would be to take a picture of periodic quantum dynamics on cubic
graphs, even if here we do not state any further general result, beyond a
crude analytic\textbf{\ }compilation of matrix entries.%

It is an open problem to prove that $P_{2}^{\times3}$ is the only connected
cubic graph with PST.

$\bigskip$

\textbf{Conjecture. }$\mathcal{G}_{D,3}^{R}=\{P_{2}^{\times3}\}$ if $D=3$ and
$\mathcal{G}_{D,3}^{R}=\emptyset$, otherwise.

\bigskip

Periodicity is not necessary for PST. There are examples of regular graphs
that are not periodic but have PST \cite{ru}.

\section{Proof of the theorem}

\subsection{Integral graphs}

The \emph{spectrum} of a graph $G$ is the multiset $\{\lambda_{1}^{[m_{1}%
]}(G),\lambda_{2}^{[m_{2}]}(G),...,\lambda_{r}^{[m_{r}]}(G)\}$ of the
eigenvalues of $A(G)$. The index $[m_{i}]$ in $\lambda_{i}^{[m_{i}]}(G)$
denotes the multiplicity of the eigenvalue $\lambda_{i}(G)$. For example, the
spectrum of the complete graph on four vertices, $K_{4}$, is $\{3,-1^{[3]}\}$.
A graph is said to be \emph{integral} if its eigenvalues are integers. There
is basically one survey on this area \cite{ba} . See also \cite{cds} for the
relevant terminology and \cite{ah} for a nontrivial upper bounds on the total
number of integral graphs with $n$ vertices. There are a few general results
that establish a relation between PST and integral graphs \cite{god}. The next
statement is directly useful to our purposes:

\begin{description}
\item[P0.] A connected regular graph is periodic if and only if it is integral
(Corollary 2.3 \cite{god}).
\end{description}

The converse is not necessarily true. This fact can be observed in the
examples discussed below. Moreover, integer eigenvalues have a weaker role in
nonregular graphs. Let us consider, for instance, a graph $W$ with eight
vertices $\{1,2,...,8\}$ and set of edges $E(W)=\{\{1,i\},\{j,8\}:2\leq
i,j\leq7;\}$ (see Fig. \ref{graphprima}). The unitary governing the dynamics
in $W$ is defined by
\[
\lbrack U_{W}\left(  t\right)  ]_{i,j}=\left\{
\begin{tabular}
[c]{rr}%
$-\sin\left(  \sqrt{3}t\right)  ^{2},$ & if $i=1$ and $j=8$ or \emph{viz.};\\
$-i\sin\left(  2\sqrt{3}t\right)  /2\sqrt{3},$ & if $\{i,j\}\in E(W)$;\\
$\cos\left(  \sqrt{3}t\right)  ^{2},$ & If $i=j=1,8$;\\
$\left(  5+\cos\left(  2\sqrt{3}t\right)  \right)  /6,$ & if $i=j\neq1,8$;\\
$-\sin\left(  \sqrt{3}t\right)  ^{2},$ & otherwise.
\end{tabular}
\ \right.
\]
There is PST between vertices $1$ and $8$, because $[U_{W}\left(  \pi
/2\sqrt{3}\right)  ]_{1,8}=-\sin\left(  3\pi/2\right)  ^{2}=-1$. At the same
time, the $6\times6$ submatrix of $U_{W}\left(  \pi/2\sqrt{3}\right)  $,
excluding the first/last row/column, is $2/3$ in the diagonal and $-1/3$
off-diagonal. The spectrum of $W$ is $\{\pm2\sqrt{3},0^{\left[  6\right]  }%
\}$. Thus, $W$ is not an integral graph.

We shall be interested in a special family of graphs. A \emph{cubic} graph is
a $3$-regular graph. All cubic integral graphs have been classified and
explicitly constructed \cite{bc, s}. In particular,

\begin{description}
\item[P1. ] There are exactly thirteen connected cubic integral graphs.
\end{description}

On the light of the statements P0 and P1, a proof of the theorem can be
obtained via a case by case analysis. As we have seen, and this is an already
known fact, the $3$-dimensional cube, $P_{2}^{\times3}$, has PST between its
antipodal vertices. These are vertices at distance three. We shall verify that
none of the remaining twelve connected, cubic integral graphs affords PST. All
graphs considered in this section are periodic.

\subsection{The complete graph $K_{4}$, the complete bipartite graph $K_{3,3}
$, and two connected copies of $K_{2,3}$}

A \emph{complete graph} on $n$ vertices is a graph $K_{n}=\left(
\{1,...,n\},E\right)  $, where $E(K_{n})=\{\{i,j\}:1\leq i,j\leq n\}$. The $ij
$-entries of $U_{K_{4}}$ are given by
\[
\lbrack U_{K_{4}}(t)]_{i,j}=\left\{
\begin{tabular}
[c]{rr}%
$\left(  3\cos\left(  t\right)  +\cos\left(  3t\right)  +4i\sin\left(
t\right)  ^{3}\right)  /4,$ & if $i=j$;\\
$\cos\left(  t\right)  \sin\left(  t\right)  \left(  -i\cos\left(  t\right)
-\sin\left(  t\right)  \right)  ,$ & if $i\neq j$.
\end{tabular}
\ \right.
\]
Then $\max_{t\in\mathbb{R}^{+}}\left(  [U_{K_{4}}(t)]_{i,i}\right)
=f_{U_{K_{4}}}\left(  i,i;\pi/2\right)  =1$, for $1\leq i\leq4$; for every
pair of vertices $i$ and $j$, $\max_{t\in\mathbb{R}^{+}}\left(  [U_{K_{4}%
}(t)]_{i,j}\right)  =f_{U_{K_{4}}}\left(  i,j;\pi/4\right)  =1/2$. Note that%
\[
\lbrack U_{K_{4}}(\pi/4)]_{i,j}=\dfrac{1}{\sqrt{2}}\left\{
\begin{tabular}
[c]{rr}%
$(1+i)/2,$ & if $i=j$;\\
$-(1+i)/2,$ & if $i\neq j$.
\end{tabular}
\ \right.
\]
gives a complex Hadamard matrix.

A graph $G=(V=V_{1}\cup V_{2},E)$ is \emph{bipartite} if each vertex in
$V_{1}$ is adjacent to vertices in $V_{2}$ only and \emph{viz}. A
\emph{complete bipartite} \emph{graph }is a bipartite graph $K_{q,p}%
=(V=V_{1}\cup V_{2},E)$ such that $\left\vert V_{1}\right\vert =p$,
$\left\vert V_{2}\right\vert =q$ and $E=\{\{i,j\}:i\in V_{1}$ and $j\in
V_{2}\}$. The graph $K_{3,3}$ is on six vertices; $V(K_{3,3})=\{A,B\}$, with
$\left\vert A\right\vert =\left\vert B\right\vert =3$. By definition,
$\{i,j\}\in E(K_{3,3})$ if and only if $i\in A$ and $j\in B$. The spectrum of
$K_{3,3}$ is $\{\pm3,0^{[4]}\}$ and dia$\left(  K_{3,3}\right)  =2$. The
$ij$-entries of $U_{K_{3,3}}$ are as follows:%
\[
\lbrack U_{K_{3,3}}(t)]_{i,j}=\left\{
\begin{tabular}
[c]{rr}%
$\left(  2+\cos\left(  3t\right)  \right)  /3,$ & $i=j;$\\
$\left(  -1+\cos\left(  3t\right)  \right)  /3,$ & $\{i,j\}\notin E(K_{3,3}%
);$\\
$-i\sin\left(  3t\right)  /3,$ & $\{i,j\}\in E(K_{3,3}).$%
\end{tabular}
\right.
\]
Hence, $\max_{t\in\mathbb{R}^{+}}\left(  \left\vert [U_{K_{3,3}}%
(t)]_{i,i}\right\vert \right)  =f_{U_{K_{3,3}}}\left(  i,i;2\pi/3\right)  =1$,
for $1\leq i\leq6$. For the off-diagonal entries, we need to distinguish two
cases: \emph{(i) }$\max_{t\in\mathbb{R}^{+}}\left(  \left\vert [U_{K_{3,3}%
}(t)]_{i,j}\right\vert \right)  =f_{U_{K_{3,3}}}\left(  i,j;\pi/3\right)
=2/3$, if $i,j\in A$ or $i,j\in B$; \emph{(ii)} $\max_{t\in\mathbb{R}^{+}%
}\left(  [\left\vert U_{K_{3,3}}(t)]_{i,j}\right\vert \right)  =f_{U_{K_{3,3}%
}}\left(  i,j;\pi/6\right)  =1/3$, if $i\in A$ and $j\in B$. When $t=\pi/2$,
\[
\lbrack U_{K_{3,3}}(\pi/2)]_{i,j}=\left\{
\begin{tabular}
[c]{rr}%
$2/3,$ & $i=j;$\\
$-1/3,$ & $\{i,j\}\notin E(K_{3,3});$\\
$i/3,$ & $\{i,j\}\in E(K_{3,3}).$%
\end{tabular}
\right.
\]

Let $DK_{2,3}$ be the graph on ten vertices obtained from two disjoint copies
of $K_{2,3}$, say $K_{2,3}^{1}$ and $K_{2,3}^{2}$, by adding three edges
between the vertices of degree two in $K_{2,3}^{1}$ and $K_{2,3}^{2}$. The
spectrum of $DK_{2,3}$ is $\{\pm3,\pm2,\pm1^{[2]},0^{[2]}\}$ and dia$\left(
DK_{2,3}\right)  =3$. The structure of $U_{DK_{3,3}}$ consists of various kind
of entries:%
\[
U_{DK_{2,3}}(t)=%
\begin{tabular}
[c]{cccccccccc}%
$a_{1}$ & $a_{2}$ & $a_{3}$ & $a_{3}$ & $a_{3}$ & $a_{4}$ & $a_{4}$ & $a_{4}$
& $a_{5}$ & $a_{5}$\\
$a_{2}$ & $a_{1}$ & $a_{3}$ & $a_{3}$ & $a_{3}$ & $a_{4}$ & $a_{4}$ & $a_{4}$
& $a_{5}$ & $a_{5}$\\\cline{3-8}\cline{3-8}%
$a_{3}$ & $a_{3}$ & \multicolumn{1}{|c}{$a_{6}$} & $a_{7}$ &
\multicolumn{1}{c|}{$a_{7}$} & $a_{8}$ & $a_{9}$ & \multicolumn{1}{c|}{$a_{9}
$} & \multicolumn{1}{|c}{$a_{4}$} & $a_{4}$\\
$a_{3}$ & $a_{3}$ & \multicolumn{1}{|c}{$a_{7}$} & $a_{6}$ &
\multicolumn{1}{c|}{$a_{7}$} & $a_{9}$ & $a_{8}$ & \multicolumn{1}{c|}{$a_{9}
$} & \multicolumn{1}{|c}{$a_{4}$} & $a_{4}$\\
$a_{3}$ & $a_{3}$ & \multicolumn{1}{|c}{$a_{7}$} & $a_{7}$ &
\multicolumn{1}{c|}{$a_{6}$} & $a_{9}$ & $a_{9}$ & \multicolumn{1}{c|}{$a_{8}
$} & \multicolumn{1}{|c}{$a_{4}$} & $a_{4}$\\\cline{3-8}\cline{3-8}%
$a_{4}$ & $a_{4}$ & \multicolumn{1}{|c}{$a_{8}$} & $a_{9}$ &
\multicolumn{1}{c|}{$a_{9}$} & \multicolumn{1}{|c}{$a_{6}$} & $a_{7}$ &
$a_{7}$ & \multicolumn{1}{|c}{$a_{3}$} & $a_{3}$\\
$a_{4}$ & $a_{4}$ & \multicolumn{1}{|c}{$a_{9}$} & $a_{8}$ &
\multicolumn{1}{c|}{$a_{9}$} & \multicolumn{1}{|c}{$a_{7}$} & $a_{6}$ &
$a_{7}$ & \multicolumn{1}{|c}{$a_{3}$} & $a_{3}$\\
$a_{4}$ & $a_{4}$ & \multicolumn{1}{|c}{$a_{9}$} & $a_{9}$ &
\multicolumn{1}{c|}{$a_{8}$} & \multicolumn{1}{|c}{$a_{7}$} & $a_{7}$ &
$a_{6}$ & \multicolumn{1}{|c}{$a_{3}$} & $a_{3}$\\\cline{3-5}\cline{3-8}%
\cline{6-8}%
$a_{5}$ & $a_{5}$ & $a_{4}$ & $a_{4}$ & $a_{4}$ & $a_{3}$ & $a_{3}$ & $a_{3}$
& $a_{1}$ & $a_{2}$\\
$a_{5}$ & $a_{5}$ & $a_{4}$ & $a_{4}$ & $a_{4}$ & $a_{3}$ & $a_{3}$ & $a_{3}$
& $a_{2}$ & $a_{1}$%
\end{tabular}
\text{ },
\]
where
\[
\begin{tabular}
[c]{l}%
$a_{1}=\left(  5+3\cos\left(  2t\right)  +2\cos\left(  3t\right)  \right)
/10,$\\
$a_{2}=\left(  -5+3\cos\left(  2t\right)  +2\cos\left(  3t\right)  \right)
/10,$\\
$a_{3}=-i\left(  \sin\left(  2t\right)  +\sin\left(  3t\right)  \right)
/5,$\\
$a_{4}=\left(  -\cos\left(  2t\right)  +\cos\left(  3t\right)  \right)  /5,$\\
$a_{5}=i\left(  3\sin\left(  2t\right)  -2\sin\left(  3t\right)  \right)
/10,$\\
$a_{6}=\left(  10\cos\left(  t\right)  +2\cos\left(  2t\right)  +3\cos\left(
3t\right)  \right)  /15,$\\
$a_{7}=\left(  -5\cos\left(  t\right)  +2\cos\left(  2t\right)  +3\cos\left(
3t\right)  \right)  /15,$\\
$a_{8}=-i\left(  10\sin\left(  t\right)  -2\sin\left(  2t\right)
+3\sin\left(  3t\right)  \right)  /15,$\\
$a_{9}=i\left(  10\sin\left(  t\right)  +2\sin\left(  2t\right)  -3\sin\left(
3t\right)  \right)  /15.$%
\end{tabular}
\]
By considering $a_{1}$ and $a_{6}$, we can see that $\max_{t\in\mathbb{R}^{+}%
}\left(  \left\vert [U_{DK_{2,3}}(t)]_{i,i}\right\vert \right)  =f_{DK_{2,3}%
}\left(  i,i;2\pi\right)  =1$, for every $i$. However, $\max_{1\leq
j\leq9;j\neq1,6}\max_{t\in\mathbb{R}^{+}}\left(  \left\vert a_{j}\right\vert
\right)  =\left(  5-5\left(  -1-\sqrt{5}\right)  /4\right)  /10\approx0.9$.
The maximum is attained by $a_{2}$ for $t=2\pi/5$. The segments in the matrix
$U_{DK_{3,3}}(t)$ help visualizing its structure and these do not have a
mathematical meaning. We shall make a consistent use of this graphic tool also
in the next cases. The graphs $K_{3,3}$ and $DK_{2,3}$ are in Fig.
(\ref{graphs1}).

\subsection{The graphs $C_{3}+K_{2}$ and $C_{6}+K_{2}$}

The (\emph{Cartesian}) \emph{sum} $G=G_{1}+G_{2}=(V,E)$ of two graphs
$G_{1}=\left(  V_{1},E_{1}\right)  $ and $G_{2}=\left(  V_{2},E_{2}\right)  $
has set of vertices $V(G)=V(G_{1})\times V(G_{2})$ and $\{\{i,j\},\{k,l\}\}\in
E\left(  G\right)  $ if \emph{(i) }$\{i,k\}\in E(G_{1})$ or \emph{(ii)
}$\{j,l\}\in E(G_{2})$ \cite{ore}. We denote by $C_{n}$ the $n$-\emph{cycle}:
$V(C_{n})=\{1,2,...,n\}$ and $\{i,\left(  i+1\right)  \operatorname{mod}n\}\in
E(C_{n})$.

The graph $C_{3}+K_{2}$ has two cycles of length three and it can be drawn as
a prism with triangular basis. It is the undirected version the Cayley digraph
of the dihedral group $D_{6}$ generated by the standard set. Its spectrum is
$\{3,1,0^{[2]},-2^{[2]}\}$. Given the symmetry, the unitary matrix has a neat
structure:%
\[
U_{C_{3}+K_{2}}(t)=%
\begin{tabular}
[c]{lll|lll}%
$a_{1}$ & $a_{2}$ & $a_{2}$ & $a_{3}$ & $a_{4}$ & $a_{4}$\\
$a_{2}$ & $a_{1}$ & $a_{2}$ & $a_{4}$ & $a_{3}$ & $a_{4}$\\
$a_{2}$ & $a_{2}$ & $a_{1}$ & $a_{4}$ & $a_{4}$ & $a_{3}$\\\hline
$a_{3}$ & $a_{4}$ & $a_{4}$ & $a_{1}$ & $a_{2}$ & $a_{2}$\\
$a_{4}$ & $a_{3}$ & $a_{4}$ & $a_{2}$ & $a_{1}$ & $a_{2}$\\
$a_{4}$ & $a_{4}$ & $a_{3}$ & $a_{2}$ & $a_{2}$ & $a_{1}$%
\end{tabular}
\text{ },
\]
with%
\[
\begin{tabular}
[c]{l}%
$a_{1}=\left(  2+e^{-it}+2e^{2it}+e^{-3it}\right)  /6,$\\
$a_{2}=-ie^{-it/2}\left(  \sin\left(  t/2\right)  +\sin\left(  5t/2\right)
\right)  /3,$\\
$a_{3}=\left(  2-e^{-it}-2e^{2it}+e^{-3it}\right)  /6,$\\
$a_{4}=\left(  -1-e^{-it}+e^{2it}+e^{-3it}\right)  /6.$%
\end{tabular}
\]
From this, $\max_{2\leq j\leq4}\max_{t\in\mathbb{R}^{+}}\left(  \left\vert
a_{j}\right\vert \right)  \approx0.9$, for $j=3$ and $t=\sqrt{3}+\pi/17$.

\begin{figure}
[ptb]
\begin{center}
\includegraphics[
trim=0.000000in 0.000000in -0.011111in 0.000000in,
height=1.0449in,
width=3.736in
]%
{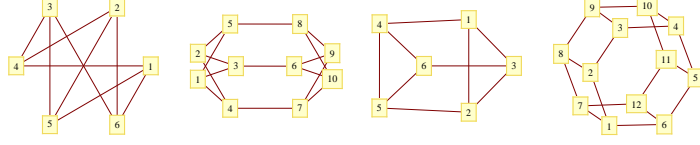}%
\caption{From the left: The complete bipartite graph $K_{3,3}$. The graph
obtained by connecting together two copies of $K_{2,3}$. The graph
$C_{3}+K_{2}$. The graph $C_{6}+K_{2}$. Although periodic, none of these
graphs has PST. }%
\label{graphs1}%
\end{center}
\end{figure}

The graph $C_{6}+K_{2}$ is on twelve vertices. It has two cycles of length six
and it can be drawn as a prism with hexagonal basis. In fact, in analogy with
$C_{3}+K_{2}$, it is the undirected version the Cayley digraph of the dihedral
group $D_{12}$ generated by the standard set. As we have done for the other
cases, we explicitly write down the unitary matrix. Let%
\[
\begin{array}
[c]{ccc}%
A=%
\begin{tabular}
[c]{lll|lll}%
$a_{1}$ & $a_{2}$ & $a_{3}$ & $a_{4}$ & $a_{3}$ & $a_{2}$\\
$a_{2}$ & $a_{1}$ & $a_{2}$ & $a_{3}$ & $a_{4}$ & $a_{3}$\\
$a_{3}$ & $a_{2}$ & $a_{1}$ & $a_{2}$ & $a_{3}$ & $a_{4}$\\\hline
$a_{4}$ & $a_{3}$ & $a_{2}$ & $a_{1}$ & $a_{2}$ & $a_{3}$\\
$a_{3}$ & $a_{4}$ & $a_{3}$ & $a_{2}$ & $a_{1}$ & $a_{2}$\\
$a_{2}$ & $a_{3}$ & $a_{4}$ & $a_{3}$ & $a_{2}$ & $a_{1}$%
\end{tabular}
& \text{and} & B=%
\begin{tabular}
[c]{lll|lll}%
$a_{5}$ & $a_{6}$ & $a_{7}$ & $a_{8}$ & $a_{7}$ & $a_{6}$\\
$a_{6}$ & $a_{5}$ & $a_{6}$ & $a_{7}$ & $a_{8}$ & $a_{7}$\\
$a_{7}$ & $a_{6}$ & $a_{5}$ & $a_{6}$ & $a_{7}$ & $a_{8}$\\\hline
$a_{8}$ & $a_{7}$ & $a_{6}$ & $a_{5}$ & $a_{6}$ & $a_{7}$\\
$a_{7}$ & $a_{8}$ & $a_{7}$ & $a_{6}$ & $a_{5}$ & $a_{6}$\\
$a_{6}$ & $a_{7}$ & $a_{8}$ & $a_{7}$ & $a_{6}$ & $a_{5}$%
\end{tabular}
\text{ }.
\end{array}
\]
Then%
\[
U_{C_{6}+K_{2}}(t)=\left[
\begin{array}
[c]{cc}%
A & B\\
B & A
\end{array}
\right]  ,
\]
and%
\[
\begin{tabular}
[c]{l}%
$a_{1}=\left(  2+\cos\left(  t\right)  +2\cos\left(  2t\right)  +\cos\left(
3t\right)  \right)  /6,$\\
$a_{2}=-i\left(  \sin\left(  t\right)  +\sin\left(  2t\right)  +\sin\left(
3t\right)  \right)  /6,$\\
$a_{3}=\left(  -1+\cos\left(  t\right)  -\cos\left(  2t\right)  +\cos\left(
3t\right)  \right)  /6,$\\
$a_{4}=i\left(  \sin\left(  t\right)  -2\sin\left(  2t\right)  +\sin\left(
3t\right)  \right)  /6,$\\
$a_{5}=i\left(  \sin\left(  t\right)  -2\sin\left(  2t\right)  -\sin\left(
3t\right)  \right)  /6,$\\
$a_{6}=-\left(  1+2\cos(t)\right)  \sin\left(  t\right)  ^{2}/3,$\\
$a_{7}=-i\left(  \sin\left(  t\right)  +\sin\left(  2t\right)  -\sin\left(
3t\right)  \right)  /6,$\\
$a_{8}=16\left(  \cos\left(  t/2\right)  ^{2}\sin\left(  t/2\right)
^{4}\right)  /3.$%
\end{tabular}
\]
For $j\neq1$,%
\[
\max_{t\in\mathbb{R}^{+}}\left(  \left\vert a_{j}\right\vert \right)  \left\{
\begin{tabular}
[c]{rr}%
$\approx21/50,$ & $j=2$ and $t\approx\pi/6;$\\
$=2/3,$ & $j=3$ and $t=\pi;$\\
$\approx29/50,$ & $j=4$ and $t=77/20$\\
$\approx9/20,$ & $j=5$ and $t=19/10;$\\
$\approx1/2,$ & $j=6$ and $t\approx1.21;$\\
$\approx9/25,$ & $j=7$ and $t\approx1.35;$\\
$\approx64/81,$ & $j=8$ and $t\approx1.9.$%
\end{tabular}
\right.
\]
The graphs $C_{3}+K_{2}$, and $C_{6}+K_{2}$ are represented in Fig.
(\ref{graphs1}).%

\subsection{The Petersen graph, a graph on ten vertices, and $L\left(
S\left(  K_{4}\right)  \right)  $}

The Petersen graph, $P$, illustrated in Fig. (\ref{peter}), is one of the
best-studied single objects in the graph-theoretic literature. The Petersen
graph has two cycles of length five. It is vertex-transitive but not a Cayley
graph, with spectrum $\{3,1^{[4]},-2^{[4]}\}$ and dia$\left(  P\right)  =3$.
It is strongly regular with parameters $(10,3,0,1)$. The symmetry appears also
in $U_{P}(t)$, which reflects faithfully the structure of $P$:
\[
U_{P}(t)=%
\begin{tabular}
[c]{lllll|lllll}%
$a_{1}$ & $a_{2}$ & $a_{3}$ & $a_{3}$ & $a_{2}$ & $a_{2}$ & $a_{3}$ & $a_{3}$
& $a_{3}$ & $a_{3}$\\
$a_{2}$ & $a_{1}$ & $a_{2}$ & $a_{3}$ & $a_{3}$ & $a_{3}$ & $a_{2}$ & $a_{3}$
& $a_{3}$ & $a_{3}$\\
$a_{3}$ & $a_{2}$ & $a_{1}$ & $a_{2}$ & $a_{3}$ & $a_{3}$ & $a_{3}$ & $a_{2}$
& $a_{3}$ & $a_{3}$\\
$a_{3}$ & $a_{3}$ & $a_{2}$ & $a_{1}$ & $a_{2}$ & $a_{3}$ & $a_{3}$ & $a_{3}$
& $a_{2}$ & $a_{3}$\\
$a_{2}$ & $a_{3}$ & $a_{3}$ & $a_{2}$ & $a_{1}$ & $a_{3}$ & $a_{3}$ & $a_{3}$
& $a_{3}$ & $a_{2}$\\\hline
$a_{2}$ & $a_{3}$ & $a_{3}$ & $a_{3}$ & $a_{3}$ & $a_{1}$ & $a_{3}$ & $a_{2}$
& $a_{2}$ & $a_{3}$\\
$a_{3}$ & $a_{2}$ & $a_{3}$ & $a_{3}$ & $a_{3}$ & $a_{3}$ & $a_{1}$ & $a_{3}$
& $a_{2}$ & $a_{2}$\\
$a_{3}$ & $a_{3}$ & $a_{2}$ & $a_{3}$ & $a_{3}$ & $a_{2}$ & $a_{3}$ & $a_{1}$
& $a_{3}$ & $a_{2}$\\
$a_{3}$ & $a_{3}$ & $a_{3}$ & $a_{2}$ & $a_{3}$ & $a_{2}$ & $a_{2}$ & $a_{3}$
& $a_{1}$ & $a_{3}$\\
$a_{3}$ & $a_{3}$ & $a_{3}$ & $a_{3}$ & $a_{2}$ & $a_{3}$ & $a_{2}$ & $a_{2}$
& $a_{3}$ & $a_{1}$%
\end{tabular}
\text{ },
\]
where%
\[
\begin{tabular}
[c]{l}%
$a_{1}=e^{-3it}\left(  1+5e^{2it}+4e^{5it}\right)  /10,$\\
$a_{2}=e^{-3it}\left(  3+5e^{2it}-8e^{5it}\right)  /30,$\\
$a_{3}=e^{-3it}\left(  3-5e^{2it}+2e^{5it}\right)  /30.$%
\end{tabular}
\]
Thus, $\max_{j=2,3}\max_{t\in\mathbb{R}^{+}}\left(  \left\vert a_{j}%
\right\vert \right)  =8/15$, for $j=2$ and $t=\pi$. More generally,
$[U_{P}(\pi)]_{i,j}=-1/5$ if $i=j$; $2/15$ if $\{i,j\}\notin E(P)$ and
$-8/15$, otherwise.

There is another cubic integral graph on ten vertices, obtained by replacing
with triangles two nonadjacent vertices of $K_{3,3}$. Denoted by $Z$, it has
spectrum $\{3,2,1^{[3]},-1^{[2]},-2^{[3]}\}$. Fig. (\ref{peter}) contains a
drawing. The unitary matrix obtained from $Z$ is%
\[
U_{Z}(t)=%
\begin{tabular}
[c]{llll|llllll}\cline{1-4}%
\multicolumn{1}{|l}{$a_{1}$} & $a_{2}$ & $a_{2}$ & $a_{2}$ & $a_{3}$ & $a_{3}
$ & $a_{3}$ & $a_{3}$ & $a_{3}$ & $a_{3}$\\
\multicolumn{1}{|l}{$a_{2}$} & $a_{1}$ & $a_{3}$ & $a_{3}$ & $a_{3}$ & $a_{3}
$ & $a_{2}$ & $a_{3}$ & $a_{2}$ & $a_{3}$\\
\multicolumn{1}{|l}{$a_{2}$} & $a_{3}$ & $a_{1}$ & $a_{3}$ & $a_{2}$ & $a_{2}
$ & $a_{3}$ & $a_{3}$ & $a_{3}$ & $a_{3}$\\
\multicolumn{1}{|l}{$a_{2}$} & $a_{3}$ & $a_{3}$ & $a_{1}$ & $a_{3}$ & $a_{3}
$ & $a_{3}$ & $a_{2}$ & $a_{3}$ & $a_{2}$\\\hline
$a_{3}$ & $a_{3}$ & $a_{2}$ & $a_{3}$ & $a_{5}$ & $a_{4}$ & $a_{6}$ & $a_{6}$
& $a_{7}$ & \multicolumn{1}{l|}{$a_{7}$}\\
$a_{3}$ & $a_{3}$ & $a_{2}$ & $a_{3}$ & $a_{4}$ & $a_{5}$ & $a_{7}$ & $a_{7}$
& $a_{6}$ & \multicolumn{1}{l|}{$a_{6}$}\\
$a_{3}$ & $a_{2}$ & $a_{3}$ & $a_{3}$ & $a_{6}$ & $a_{7}$ & $a_{5}$ & $a_{6}$
& $a_{4}$ & \multicolumn{1}{l|}{$a_{7}$}\\
$a_{3}$ & $a_{3}$ & $a_{3}$ & $a_{2}$ & $a_{6}$ & $a_{7}$ & $a_{6}$ & $a_{5}$
& $a_{7}$ & \multicolumn{1}{l|}{$a_{4}$}\\
$a_{3}$ & $a_{2}$ & $a_{3}$ & $a_{3}$ & $a_{7}$ & $a_{6}$ & $a_{4}$ & $a_{7}$
& $a_{5}$ & \multicolumn{1}{l|}{$a_{6}$}\\
$a_{3}$ & $a_{3}$ & $a_{3}$ & $a_{2}$ & $a_{7}$ & $a_{6}$ & $a_{7}$ & $a_{4}$
& $a_{6}$ & \multicolumn{1}{l|}{$a_{5}$}\\\cline{5-10}%
\end{tabular}
\text{ .}
\]
The entries are%
\[
\begin{tabular}
[c]{l}%
$a_{1}=e^{-3it}\left(  1+5e^{2it}+4e^{5it}\right)  /10,$\\
$a_{2}=e^{-3it}\left(  3+5e^{2it}-8e^{5it}\right)  /30,$\\
$a_{3}=e^{-3it}\left(  3-5e^{2it}+2^{5it}\right)  /30,$\\
$a_{4}=e^{-3it}\left(  e^{it}-1\right)  ^{2}\left(  3+e^{it}+4e^{2it}%
+7e^{3it}\right)  /30,$\\
$a_{5}=e^{-3it}\left(  3+5e^{it}+5e^{2it}+10e^{4it}+7e^{5it}\right)  /30,$\\
$a_{6}=e^{-3it}\left(  3+5e^{it}-5e^{4it}-3e^{5it}\right)  /30,$\\
$a_{7}=e^{-3it}\left(  3-5e^{it}+5e^{4it}-3^{5it}\right)  /30.$%
\end{tabular}
\]
We then obtain $\max_{1\leq j\leq9;j\neq1,5}\max_{t\in\mathbb{R}^{+}}\left(
\left\vert a_{j}\right\vert \right)  \approx0.85$, for $j=4$ and $t\approx
\pi-5/6$. Note that $a_{1}$, $a_{2}$, and $a_{3}$ give the same dynamics as
the Petersen graph.

The graph $X=L\left(  S\left(  K_{4}\right)  \right)  $ is constructed by
replacing each vertex of $K_{4}$ with a triangle (see Fig. (\ref{peter})). The
triangles are then connected by independent edges. The notation indicates the
line graph of the $K_{4}$ subdivision. Its spectrum is $\{3,\pm2^{[3]}%
,0^{[2]},-1^{[3]}\}$ and dia$\left(  X\right)  =3$. The unitary has some
symmetry:%
\[
U_{X}(t)=%
\begin{tabular}
[c]{llllllllllll}\cline{1-3}\cline{8-8}%
\multicolumn{1}{|l}{$a_{1}$} & $a_{2}$ & \multicolumn{1}{l|}{$a_{2}$} &
$a_{4}$ & $a_{5}$ & $a_{6}$ & $a_{4}$ & \multicolumn{1}{|l}{$a_{3}$} &
\multicolumn{1}{|l}{$a_{4}$} & $a_{4}$ & $a_{6}$ & $a_{5}$\\\cline{4-4}%
\cline{8-8}%
\multicolumn{1}{|l}{$a_{2}$} & $a_{1}$ & \multicolumn{1}{l|}{$a_{2}$} &
\multicolumn{1}{l|}{$a_{3}$} & $a_{4}$ & $a_{4}$ & $a_{5}$ & $a_{4}$ & $a_{6}
$ & $a_{4}$ & $a_{5}$ & $a_{6}$\\\cline{4-4}\cline{10-10}%
\multicolumn{1}{|l}{$a_{2}$} & $a_{2}$ & \multicolumn{1}{l|}{$a_{1}$} &
$a_{4}$ & $a_{6}$ & $a_{5}$ & $a_{4}$ & $a_{4}$ & $a_{5}$ &
\multicolumn{1}{|l}{$a_{3}$} & \multicolumn{1}{|l}{$a_{4}$} & $a_{4}%
$\\\cline{1-6}\cline{2-2}\cline{10-10}%
$a_{4}$ & \multicolumn{1}{|l}{$a_{3}$} & \multicolumn{1}{|l}{$a_{4}$} &
\multicolumn{1}{|l}{$a_{1}$} & $a_{2}$ & \multicolumn{1}{l|}{$a_{2}$} &
$a_{4}$ & $a_{5}$ & $a_{6}$ & $a_{5}$ & $a_{4}$ & $a_{6}$\\\cline{2-2}%
\cline{7-7}%
$a_{5}$ & $a_{4}$ & $a_{6}$ & \multicolumn{1}{|l}{$a_{2}$} & $a_{1}$ &
\multicolumn{1}{l|}{$a_{2}$} & \multicolumn{1}{l|}{$a_{3}$} & $a_{4}$ &
$a_{4}$ & $a_{6}$ & $a_{4}$ & $a_{5}$\\\cline{7-7}\cline{11-11}%
$a_{6}$ & $a_{4}$ & $a_{5}$ & \multicolumn{1}{|l}{$a_{2}$} & $a_{2}$ &
\multicolumn{1}{l|}{$a_{1}$} & $a_{4}$ & $a_{6}$ & $a_{5}$ & $a_{4}$ &
\multicolumn{1}{|l}{$a_{3}$} & \multicolumn{1}{|l}{$a_{4}$}\\\cline{4-9}%
\cline{5-5}\cline{11-11}%
$a_{4}$ & $a_{5}$ & $a_{6}$ & $a_{4}$ & \multicolumn{1}{|l}{$a_{3}$} &
\multicolumn{1}{|l}{$a_{4}$} & \multicolumn{1}{|l}{$a_{1}$} & $a_{2}$ &
\multicolumn{1}{l|}{$a_{2}$} & $a_{6}$ & $a_{5}$ & $a_{4}$\\\cline{1-1}%
\cline{5-5}%
\multicolumn{1}{|l}{$a_{3}$} & \multicolumn{1}{|l}{$a_{4}$} & $a_{4}$ &
$a_{5}$ & $a_{4}$ & $a_{6}$ & \multicolumn{1}{|l}{$a_{2}$} & $a_{1}$ &
\multicolumn{1}{l|}{$a_{2}$} & $a_{5}$ & $a_{6}$ & $a_{4}$\\\cline{1-1}%
\cline{12-12}%
$a_{4}$ & $a_{6}$ & $a_{5}$ & $a_{6}$ & $a_{4}$ & $a_{5}$ &
\multicolumn{1}{|l}{$a_{2}$} & $a_{2}$ & \multicolumn{1}{l|}{$a_{1}$} &
$a_{4}$ & $a_{4}$ & \multicolumn{1}{|l|}{$a_{3}$}\\\cline{3-3}\cline{7-12}%
\cline{12-12}%
$a_{4}$ & $a_{4}$ & \multicolumn{1}{|l}{$a_{3}$} & \multicolumn{1}{|l}{$a_{5}
$} & $a_{6}$ & $a_{4}$ & $a_{6}$ & $a_{5}$ & $a_{4}$ &
\multicolumn{1}{|l}{$a_{1}$} & $a_{2}$ & \multicolumn{1}{l|}{$a_{2}$%
}\\\cline{3-3}\cline{6-6}%
$a_{6}$ & $a_{5}$ & $a_{4}$ & $a_{4}$ & $a_{4}$ & \multicolumn{1}{|l}{$a_{3}$}
& \multicolumn{1}{|l}{$a_{5}$} & $a_{6}$ & $a_{4}$ &
\multicolumn{1}{|l}{$a_{2}$} & $a_{1}$ & \multicolumn{1}{l|}{$a_{2}$%
}\\\cline{6-6}\cline{9-9}%
$a_{5}$ & $a_{6}$ & $a_{4}$ & $a_{6}$ & $a_{5}$ & $a_{4}$ & $a_{4}$ & $a_{4}$
& \multicolumn{1}{|l}{$a_{3}$} & \multicolumn{1}{|l}{$a_{2}$} & $a_{2}$ &
\multicolumn{1}{l|}{$a_{1}$}\\\cline{9-9}\cline{9-12}%
\end{tabular}
\ ,
\]
with%
\[
\begin{tabular}
[c]{l}%
$a_{1}=\left(  2+3e^{it}+3e^{-2it}+3e^{2it}+e^{-3it}\right)  /12,$\\
$a_{2}=-e^{-3it}\left(  -2-5e^{it}+2e^{3it}+2e^{4it}+3e^{5it}\right)  /24,$\\
$a_{3}=e^{-3it}\left(  1+e^{it}+2e^{3it}-e^{4it}-e^{5it}\right)  /12,$\\
$a_{4}=e^{-3it}\left(  e^{it}-1\right)  ^{2}\left(  2+3e^{it}+4e^{2it}%
+3e^{3it}\right)  /24,$\\
$a_{5}=-e^{-3it}\left(  e^{it}-1\right)  ^{3}\left(  2+3e^{it}+3e^{2it}%
\right)  /24,$\\
$a_{6}=-e^{-3it}\left(  e^{it}-1\right)  ^{3}\left(  e^{it}+1\right)  /12.$%
\end{tabular}
\
\]
For $j\neq1$,%
\[
\max_{t\in\mathbb{R}^{+}}\left(  \left\vert a_{j}\right\vert \right)  \left\{
\begin{tabular}
[c]{rr}%
$\approx1/2,$ & $j=2$ and $t\approx\pi/4;$\\
$\approx1/2,$ & $j=3$ and $t\approx\pi/4;$\\
$\approx1/4,$ & $j=4$ and $t\approx6/5;$\\
$=2/3,$ & $j=5$ and $t=\pi;$\\
$=\sqrt{3}/4,$ & $j=6$ and $t=2\pi/3.$%
\end{tabular}
\ \right.
\]

\textbf{%
\begin{figure}
[ptb]
\begin{center}
\includegraphics[
height=1.1344in,
width=3.6322in
]%
{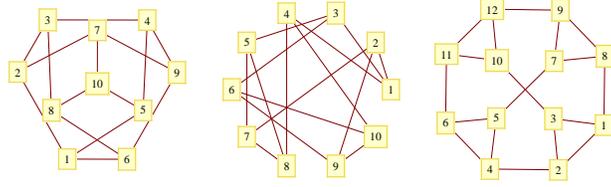}%
\caption{\emph{(L)}: The Petersen graph. \emph{(C)}: The graph $Z$ on ten
vertices. \emph{(R)}: The graph $L\left(  S\left(  K_{4}\right)  \right)  $ on
twelve vertices. These graphs are periodic without PST. }%
\label{peter}%
\end{center}
\end{figure}
}

\subsection{The Desargues graph and its cospectral mate}

\begin{figure}
[ptb]
\begin{center}
\includegraphics[
height=1.8263in,
width=3.4823in
]%
{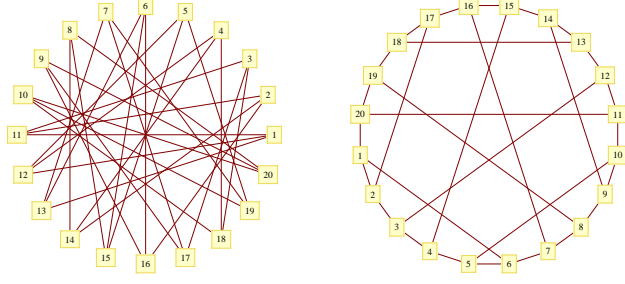}%
\caption{\emph{(L):} The graph $H_{5,2}^{\prime}$. \emph{(R):} The Desargues
graph $H_{5,2}$. This is cospectral with $H_{5,2}^{\prime}$. }%
\label{des}%
\end{center}
\end{figure}

The bipartite double cover of the Petersen graph is called \emph{Desargues
graph}. There are many different notations for this graph. We adopt $H_{5,2}$.
The Desargues graph is on twenty vertices and its spectrum is $\left\{
\pm3,\pm2^{[4]},\pm1^{[5]}\right\}  $. The graph $H_{5,2}$ has a
\emph{cospectral mate}, which we will denote by $H_{5,2}^{\prime}$. This is a
nonisomorphic graph with the same spectrum. Two graphs $G$ and $H$ are said to
\emph{isomorphic} if there is a permutation matrix $Q$ such that
$QA(G)Q^{T}=A(H)$. It is clear that two isomorphic graphs have the same
spectrum. The converse is not necessarily true. Indeed, $H_{5,2}$ and
$H_{5,2}^{\prime}$ are a counterexample. The graphs $H_{5,2}$ and
$H_{5,2}^{\prime}$ are drawn in Fig. (\ref{des}). Let us define the arrays%
\[
A=%
\begin{tabular}
[c]{llllllllll}%
$a_{1}$ & $a_{2}$ & $a_{2}$ & $a_{2}$ & $a_{3}$ & $a_{2}$ & $a_{2}$ & $a_{2}$
& $a_{3}$ & $a_{3}$\\
$a_{2}$ & $a_{1}$ & $a_{2}$ & $a_{2}$ & $a_{2}$ & $a_{3}$ & $a_{3}$ & $a_{2}$
& $a_{2}$ & $a_{3}$\\
$a_{2}$ & $a_{2}$ & $a_{1}$ & $a_{3}$ & $a_{2}$ & $a_{2}$ & $a_{2}$ & $a_{3}$
& $a_{2}$ & $a_{3}$\\
$a_{2}$ & $a_{2}$ & $a_{3}$ & $a_{1}$ & $a_{2}$ & $a_{2}$ & $a_{3}$ & $a_{2}$
& $a_{3}$ & $a_{2}$\\
$a_{3}$ & $a_{2}$ & $a_{2}$ & $a_{2}$ & $a_{1}$ & $a_{2}$ & $a_{3}$ & $a_{3}$
& $a_{2}$ & $a_{2}$\\
$a_{2}$ & $a_{3}$ & $a_{2}$ & $a_{2}$ & $a_{2}$ & $a_{1}$ & $a_{2}$ & $a_{3}$
& $a_{3}$ & $a_{2}$\\
$a_{2}$ & $a_{3}$ & $a_{2}$ & $a_{3}$ & $a_{3}$ & $a_{2}$ & $a_{1}$ & $a_{2}$
& $a_{2}$ & $a_{2}$\\
$a_{2}$ & $a_{2}$ & $a_{3}$ & $a_{2}$ & $a_{3}$ & $a_{3}$ & $a_{2}$ & $a_{1}$
& $a_{2}$ & $a_{2}$\\
$a_{3}$ & $a_{2}$ & $a_{2}$ & $a_{3}$ & $a_{2}$ & $a_{3}$ & $a_{2}$ & $a_{2}$
& $a_{1}$ & $a_{2}$\\
$a_{3}$ & $a_{3}$ & $a_{3}$ & $a_{2}$ & $a_{2}$ & $a_{2}$ & $a_{2}$ & $a_{2}$
& $a_{2}$ & $a_{1}$%
\end{tabular}
,
\]%
\[
B=%
\begin{tabular}
[c]{llllllllll}%
$a_{4}$ & $a_{4}$ & $a_{4}$ & $a_{8}$ & $a_{8}$ & $a_{8}$ & $a_{8}$ & $a_{8}$
& $a_{8}$ & $a_{7}$\\
$a_{4}$ & $a_{8}$ & $a_{8}$ & $a_{4}$ & $a_{8}$ & $a_{4}$ & $a_{8}$ & $a_{8}$
& $a_{7}$ & $a_{8}$\\
$a_{4}$ & $a_{8}$ & $a_{8}$ & $a_{8}$ & $a_{7}$ & $a_{8}$ & $a_{4}$ & $a_{4}$
& $a_{8}$ & $a_{8}$\\
$a_{5}$ & $a_{4}$ & $a_{6}$ & $a_{4}$ & $a_{8}$ & $a_{6}$ & $a_{6}$ & $a_{4}$
& $a_{8}$ & $a_{8}$\\
$a_{6}$ & $a_{4}$ & $a_{5}$ & $a_{8}$ & $a_{4}$ & $a_{6}$ & $a_{6}$ & $a_{8}$
& $a_{4}$ & $a_{8}$\\
$a_{8}$ & $a_{8}$ & $a_{4}$ & $a_{8}$ & $a_{4}$ & $a_{4}$ & $a_{8}$ & $a_{7}$
& $a_{8}$ & $a_{8}$\\
$a_{8}$ & $a_{8}$ & $a_{4}$ & $a_{7}$ & $a_{8}$ & $a_{8}$ & $a_{4}$ & $a_{8}$
& $a_{4}$ & $a_{8}$\\
$a_{6}$ & $a_{8}$ & $a_{6}$ & $a_{4}$ & $a_{4}$ & $a_{5}$ & $a_{6}$ & $a_{8}$
& $a_{8}$ & $a_{4}$\\
$a_{8}$ & $a_{7}$ & $a_{8}$ & $a_{8}$ & $a_{8}$ & $a_{4}$ & $a_{4}$ & $a_{8}$
& $a_{8}$ & $a_{4}$\\
$a_{6}$ & $a_{8}$ & $a_{6}$ & $a_{8}$ & $a_{8}$ & $a_{6}$ & $a_{5}$ & $a_{4}$
& $a_{4}$ & $a_{4}$%
\end{tabular}
.
\]
Let $\widetilde{M}$ be the matrix obtained by permuting the lines (rows and
columns) of a square matrix $M$ such that line $i$ in $M$ is line $n-j+1$ in
$\widetilde{M}$. One can observe that%
\[
U_{H_{5,2}^{\prime}}(t)=\left[
\begin{array}
[c]{cc}%
A & \widetilde{B}\\
B & \widetilde{A}%
\end{array}
\right]  ,
\]
where%
\[
\begin{tabular}
[c]{l}%
$a_{1}=e^{-3it}\left(  1+4e^{it}+5e^{2it}+5e^{4it}+4e^{5it}+e^{6it}\right)
/20,$\\
$a_{2}=e^{-3it}\left(  3+2e^{it}-5e^{2it}-5e^{4it}+2e^{5it}+3e^{6it}\right)
/60,$\\
$a_{3}=e^{-3it}\left(  e^{it}-1\right)  ^{4}\left(  3+4e^{it}+3e^{2it}\right)
/60,$\\
$a_{4}=e^{-3it}\left(  3+8e^{it}+5e^{2it}-5e^{4it}-8e^{5it}-3e^{6it}\right)
/60,$\\
$a_{5}=-e^{-3it}\left(  e^{it}-1\right)  ^{3}\left(  1+4e^{it}+4e^{2it}%
+e^{3it}\right)  /20,$\\
$a_{6}=-e^{-3it}\left(  e^{it}-1\right)  ^{3}\left(  3+2e^{it}+2e^{2it}%
+3e^{3it}\right)  /60,$\\
$a_{7}=-e^{-3it}\left(  e^{it}-1\right)  ^{5}\left(  1+e^{it}\right)  /20,$\\
$a_{8}=-e^{-3it}\left(  e^{it}-1\right)  ^{3}\left(  3+7e^{it}+7e^{2it}%
+3e^{3it}\right)  /60.$%
\end{tabular}
\]
From these functions, we can see that $\max_{1\leq j\leq8;j\neq1}\max
_{t\in\mathbb{R}^{+}}\left(  \left\vert a_{j}\right\vert \right)  \approx
0.83$, for $j=7$ and $t\approx575/250$. When $t=\pi$, the probability
amplitude is supported by all vertices in a class of the bipartition; in other
words, the matrix $U_{H_{5,2}^{\prime}}(\pi)$ is block-diagonal with two
$10\times10$ blocks corresponding to the classes. The functions in the above
equations completely specify the dynamics in $H_{5,2}$, since $H_{5,2}%
^{\prime}$ and $H_{5,2}^{\prime}$ are cospectral. One can verify that the
matrix $U_{H_{5,2}}(t)$ is obtained by rearranging the entries of
$U_{H_{5,2}^{\prime}}(t)$.%

\subsection{The Nauru graph and the Tutte-Coxeter graph}

The \emph{Nauru graph}, $N_{24}$, is the only cubic symmetric (arc-transitive)
graph on $24$ vertices. It is bipartite, with spectrum $\left\{  \pm
3,\pm2^{[6]},\pm1^{\left[  3\right]  },0^{\left[  4\right]  }\right\}  $ and
dia$\left(  N_{24}\right)  =4$. A brief parenthesis: the Foster census of
cubic symmetric graphs highlights that a large portion of periodic cubic graph
is symmetric \cite{con}. Clearly, the two sets do not coincide. There are
exactly seven different kind of entries in $U_{N_{24}}(t) $. This can be seen
as a $2\times2$ block matrix. The blocks $\left(  1,1\right)  $ and $\left(
2,1\right)  $ are below. The other blocks are just their rearrangements:%
\[
\begin{tabular}
[c]{llllll|llllll}%
$a_{1}$ & $a_{2}$ & $a_{2}$ & $a_{2}$ & $a_{3}$ & $a_{2}$ & $a_{3}$ & $a_{4}$
& $a_{4}$ & $a_{3}$ & $a_{2}$ & $a_{2}$\\
$a_{2}$ & $a_{1}$ & $a_{2}$ & $a_{2}$ & $a_{2}$ & $a_{3}$ & $a_{4}$ & $a_{2}$
& $a_{3}$ & $a_{2}$ & $a_{3}$ & $a_{4}$\\
$a_{2}$ & $a_{2}$ & $a_{1}$ & $a_{3}$ & $a_{4}$ & $a_{2}$ & $a_{2}$ & $a_{3}$
& $a_{2}$ & $a_{2}$ & $a_{4}$ & $a_{3}$\\
$a_{2}$ & $a_{2}$ & $a_{3}$ & $a_{1}$ & $a_{2}$ & $a_{4}$ & $a_{2}$ & $a_{3}$
& $a_{2}$ & $a_{4}$ & $a_{2}$ & $a_{3}$\\
$a_{3}$ & $a_{2}$ & $a_{4}$ & $a_{2}$ & $a_{1}$ & $a_{2}$ & $a_{3}$ & $a_{2}$
& $a_{2}$ & $a_{3}$ & $a_{4}$ & $a_{2}$\\
$a_{2}$ & $a_{3}$ & $a_{2}$ & $a_{4}$ & $a_{2}$ & $a_{1}$ & $a_{2}$ & $a_{2}$
& $a_{3}$ & $a_{4}$ & $a_{3}$ & $a_{2}$\\\hline
$a_{3}$ & $a_{4}$ & $a_{2}$ & $a_{2}$ & $a_{3}$ & $a_{2}$ & $a_{1}$ & $a_{2}$
& $a_{2}$ & $a_{3}$ & $a_{2}$ & $a_{4}$\\
$a_{4}$ & $a_{2}$ & $a_{3}$ & $a_{3}$ & $a_{2}$ & $a_{2}$ & $a_{2}$ & $a_{1}$
& $a_{4}$ & $a_{2}$ & $a_{2}$ & $a_{3}$\\
$a_{4}$ & $a_{3}$ & $a_{2}$ & $a_{2}$ & $a_{2}$ & $a_{3}$ & $a_{2}$ & $a_{4}$
& $a_{1}$ & $a_{2}$ & $a_{3}$ & $a_{2}$\\
$a_{3}$ & $a_{2}$ & $a_{2}$ & $a_{4}$ & $a_{3}$ & $a_{4}$ & $a_{3}$ & $a_{2}$
& $a_{2}$ & $a_{1}$ & $a_{2}$ & $a_{2}$\\
$a_{2}$ & $a_{3}$ & $a_{4}$ & $a_{2}$ & $a_{4}$ & $a_{3}$ & $a_{2}$ & $a_{2}$
& $a_{3}$ & $a_{2}$ & $a_{1}$ & $a_{2}$\\
$a_{2}$ & $a_{4}$ & $a_{3}$ & $a_{3}$ & $a_{2}$ & $a_{2}$ & $a_{4}$ & $a_{3}$
& $a_{2}$ & $a_{2}$ & $a_{2}$ & $a_{1}$%
\end{tabular}
\]
and%
\[
\begin{tabular}
[c]{llllll|llllll}%
$a_{5}$ & $a_{5}$ & $a_{6}$ & $a_{5}$ & $a_{6}$ & $a_{7}$ & $a_{7}$ & $a_{7}$
& $a_{7}$ & $a_{7}$ & $a_{6}$ & $a_{7}$\\
$a_{5}$ & $a_{6}$ & $a_{5}$ & $a_{7}$ & $a_{7}$ & $a_{5}$ & $a_{6}$ & $a_{7}$
& $a_{7}$ & $a_{7}$ & $a_{7}$ & $a_{6}$\\
$a_{6}$ & $a_{7}$ & $a_{7}$ & $a_{7}$ & $a_{5}$ & $a_{5}$ & $a_{7}$ & $a_{6}$
& $a_{6}$ & $a_{7}$ & $a_{7}$ & $a_{5}$\\
$a_{7}$ & $a_{5}$ & $a_{7}$ & $a_{6}$ & $a_{5}$ & $a_{6}$ & $a_{7}$ & $a_{5}$
& $a_{7}$ & $a_{6}$ & $a_{7}$ & $a_{7}$\\
$a_{7}$ & $a_{7}$ & $a_{6}$ & $a_{7}$ & $a_{6}$ & $a_{5}$ & $a_{5}$ & $a_{5}$
& $a_{7}$ & $a_{7}$ & $a_{6}$ & $a_{7}$\\
$a_{7}$ & $a_{7}$ & $a_{5}$ & $a_{6}$ & $a_{7}$ & $a_{6}$ & $a_{5}$ & $a_{7}$
& $a_{5}$ & $a_{6}$ & $a_{7}$ & $a_{7}$\\\hline
$a_{6}$ & $a_{5}$ & $a_{5}$ & $a_{7}$ & $a_{7}$ & $a_{7}$ & $a_{7}$ & $a_{6}$
& $a_{6}$ & $a_{5}$ & $a_{7}$ & $a_{7}$\\
$a_{7}$ & $a_{7}$ & $a_{6}$ & $a_{7}$ & $a_{6}$ & $a_{7}$ & $a_{7}$ & $a_{7}$
& $a_{5}$ & $a_{5}$ & $a_{6}$ & $a_{5}$\\
$a_{7}$ & $a_{5}$ & $a_{5}$ & $a_{7}$ & $a_{7}$ & $a_{7}$ & $a_{6}$ & $a_{5}$
& $a_{7}$ & $a_{5}$ & $a_{5}$ & $a_{6}$\\
$a_{5}$ & $a_{6}$ & $a_{5}$ & $a_{6}$ & $a_{7}$ & $a_{6}$ & $a_{7}$ & $a_{7}$
& $a_{7}$ & $a_{6}$ & $a_{5}$ & $a_{5}$\\
$a_{6}$ & $a_{7}$ & $a_{5}$ & $a_{5}$ & $a_{7}$ & $a_{7}$ & $a_{5}$ & $a_{6}$
& $a_{6}$ & $a_{7}$ & $a_{5}$ & $a_{7}$\\
$a_{7}$ & $a_{7}$ & $a_{5}$ & $a_{5}$ & $a_{5}$ & $a_{7}$ & $a_{6}$ & $a_{7}$
& $a_{5}$ & $a_{7}$ & $a_{7}$ & $a_{6}$%
\end{tabular}
\text{ },
\]
where%
\[
\begin{tabular}
[c]{l}%
$a_{1}=\left(  2+3\cos\left(  t\right)  +6\cos\left(  2t\right)  +\cos\left(
3t\right)  \right)  /12,$\\
$a_{2}=-\left(  \left(  1+2\cos\left(  t\right)  \right)  \sin\left(
t\right)  ^{2}\right)  /6,$\\
$a_{3}=8\left(  \cos\left(  t/2\right)  ^{2}\sin\left(  t/2\right)
^{4}\right)  /3,$\\
$a_{4}=2\left(  \left(  1+2\cos\left(  t\right)  \right)  \sin\left(
t/2\right)  ^{4}\right)  /3,$\\
$a_{5}=-i\left(  \sin\left(  t\right)  +4\sin\left(  2t\right)  +\sin\left(
3t\right)  \right)  /12,$\\
$a_{6}=i\sin\left(  t\right)  ^{3}/3,$\\
$a_{7}=-i\left(  \sin\left(  t\right)  -2\sin\left(  2t\right)  +\sin\left(
3t\right)  \right)  /12$%
\end{tabular}
\]
are all the the different entries of the unitary matrix. From these functions,
we can write%
\[
\max_{t\in\mathbb{R}^{+}}\left(  \left\vert a_{j}\right\vert \right)
=\left\{
\begin{tabular}
[c]{rr}%
$=1,$ & $j=1$ and $t=2\pi;$\\
$\approx1/2,$ & $j=2$ and $t\approx\pi/4;$\\
$\approx1/4,$ & $j=3$ and $t\approx6/5;$\\
$=2/3,$ & $j=4$ and $t=\pi;$\\
$\approx9/20,$ & $j=5$ and $t\approx\pi/4;$\\
$=1/3,$ & $j=6$ and $t=\pi/2;$\\
$\approx3/10,$ & $j=7$ and $t\approx3\pi/4.$%
\end{tabular}
\right.
\]

\begin{figure}
[ptb]
\begin{center}
\includegraphics[
height=1.8573in,
width=3.5382in
]%
{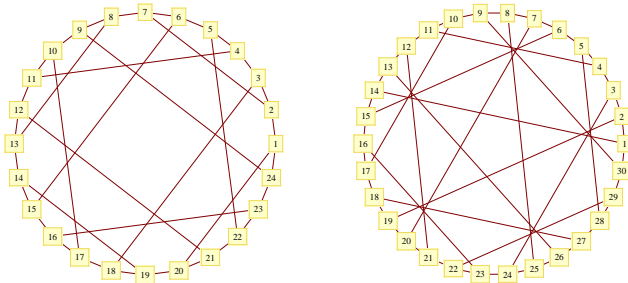}%
\caption{\emph{(L)}: The Nauru graph. \emph{(R)}: The Tutte-Coxeter graph.
This is the largest periodic cubic graph. The Nauru graph and the
Tutte-Coxeter graph do not have PST. }%
\label{tutte}%
\end{center}
\end{figure}

The \emph{Tutte-Coxeter graph}, $TC$, is illustrated in Fig. (\ref{tutte}).
This is the largest cubic graph giving a periodic dynamics. Its eigenvalues
are $\left\{  \pm3,\pm2^{\left[  9\right]  },0^{\left[  10\right]  }\right\}
$. Let us describe how to write down the unitary matrix $U_{TC}$. In Fig.
(\ref{tutte}) we have ordered the vertices anticlockwise. Each even vertex is
connected to odd vertices, and \emph{viz. }As we have seen in previous
examples, $TC$ is bipartite and its unitary has a block structure:%
\[
U_{TC}=\left[
\begin{array}
[c]{cc}%
A & B\\
B & C
\end{array}
\right]  .
\]
The $ij$-th entry of the block $B$ is%
\[
\left[  B\right]  _{i,j}=\left\{
\begin{tabular}
[c]{rr}%
$-i\left(  6\sin\left(  2t\right)  +\sin\left(  3t\right)  \right)  /15,$ &
$\{i,j\}\in E(TC);$\\
$i\left(  3\sin\left(  2t\right)  -2\sin\left(  3t\right)  \right)  /30,$ &
$\{i,j\}\notin E(TC).$%
\end{tabular}
\right.
\]
The index $i$ runs over the odd numbers $1,3,...,29$; $j$ over the even ones,
$2,4,...,30$. For any chosen $t$, none of the entries of $B$ can have unit
absolute value. The blocks $A$ and $C$ have the same entries but rearranged:%
\[
\begin{tabular}
[c]{lllllllllllllll}%
$a_{1}$ & $a_{2}$ & $a_{3}$ & $a_{3}$ & $a_{2}$ & $a_{3}$ & $a_{2}$ & $a_{2}$
& $a_{3}$ & $a_{2}$ & $a_{3}$ & $a_{3}$ & $a_{3}$ & $a_{3}$ & $a_{2}$\\
$a_{2}$ & $a_{1}$ & $a_{2}$ & $a_{3}$ & $a_{3}$ & $a_{2}$ & $a_{3}$ & $a_{3}$
& $a_{3}$ & $a_{2}$ & $a_{3}$ & $a_{2}$ & $a_{2}$ & $a_{3}$ & $a_{3}$\\
$a_{3}$ & $a_{2}$ & $a_{1}$ & $a_{2}$ & $a_{3}$ & $a_{2}$ & $a_{3}$ & $a_{2}$
& $a_{3}$ & $a_{3}$ & $a_{3}$ & $a_{3}$ & $a_{3}$ & $a_{2}$ & $a_{2}$\\
$a_{3}$ & $a_{3}$ & $a_{2}$ & $a_{1}$ & $a_{2}$ & $a_{3}$ & $a_{3}$ & $a_{2}$
& $a_{3}$ & $a_{2}$ & $a_{2}$ & $a_{3}$ & $a_{2}$ & $a_{3}$ & $a_{3}$\\
$a_{2}$ & $a_{3}$ & $a_{3}$ & $a_{2}$ & $a_{1}$ & $a_{2}$ & $a_{3}$ & $a_{3}$
& $a_{2}$ & $a_{3}$ & $a_{3}$ & $a_{3}$ & $a_{2}$ & $a_{3}$ & $a_{2}$\\
$a_{3}$ & $a_{2}$ & $a_{2}$ & $a_{3}$ & $a_{2}$ & $a_{1}$ & $a_{2}$ & $a_{3}$
& $a_{2}$ & $a_{3}$ & $a_{2}$ & $a_{3}$ & $a_{3}$ & $a_{3}$ & $a_{3}$\\
$a_{2}$ & $a_{3}$ & $a_{3}$ & $a_{3}$ & $a_{3}$ & $a_{2}$ & $a_{1}$ & $a_{2}$
& $a_{3}$ & $a_{3}$ & $a_{2}$ & $a_{3}$ & $a_{2}$ & $a_{3}$ & $a_{3}$\\
$a_{2}$ & $a_{3}$ & $a_{2}$ & $a_{2}$ & $a_{3}$ & $a_{3}$ & $a_{2}$ & $a_{1}$
& $a_{2}$ & $a_{3}$ & $a_{3}$ & $a_{2}$ & $a_{3}$ & $a_{2}$ & $a_{3}$\\
$a_{3}$ & $a_{3}$ & $a_{3}$ & $a_{3}$ & $a_{2}$ & $a_{2}$ & $a_{3}$ & $a_{2}$
& $a_{1}$ & $a_{2}$ & $a_{3}$ & $a_{2}$ & $a_{3}$ & $a_{3}$ & $a_{3}$\\
$a_{2}$ & $a_{2}$ & $a_{3}$ & $a_{2}$ & $a_{3}$ & $a_{3}$ & $a_{3}$ & $a_{3}$
& $a_{2}$ & $a_{1}$ & $a_{2}$ & $a_{3}$ & $a_{3}$ & $a_{2}$ & $a_{3}$\\
$a_{3}$ & $a_{3}$ & $a_{3}$ & $a_{2}$ & $a_{3}$ & $a_{2}$ & $a_{2}$ & $a_{3}$
& $a_{3}$ & $a_{2}$ & $a_{1}$ & $a_{2}$ & $a_{3}$ & $a_{2}$ & $a_{2}$\\
$a_{3}$ & $a_{2}$ & $a_{3}$ & $a_{3}$ & $a_{3}$ & $a_{3}$ & $a_{3}$ & $a_{2}$
& $a_{2}$ & $a_{3}$ & $a_{2}$ & $a_{1}$ & $a_{2}$ & $a_{3}$ & $a_{2}$\\
$a_{3}$ & $a_{2}$ & $a_{3}$ & $a_{2}$ & $a_{2}$ & $a_{3}$ & $a_{2}$ & $a_{3}$
& $a_{3}$ & $a_{3}$ & $a_{3}$ & $a_{2}$ & $a_{1}$ & $a_{3}$ & $a_{3}$\\
$a_{3}$ & $a_{3}$ & $a_{2}$ & $a_{3}$ & $a_{3}$ & $a_{3}$ & $a_{2}$ & $a_{3}$
& $a_{2}$ & $a_{2}$ & $a_{3}$ & $a_{3}$ & $a_{2}$ & $a_{1}$ & $a_{2}$\\
$a_{2}$ & $a_{3}$ & $a_{2}$ & $a_{3}$ & $a_{2}$ & $a_{3}$ & $a_{3}$ & $a_{3}$
& $a_{3}$ & $a_{3}$ & $a_{2}$ & $a_{2}$ & $a_{3}$ & $a_{2}$ & $a_{1}$%
\end{tabular}
,
\]
where%
\[
\begin{tabular}
[c]{l}%
$a_{1}=\left(  5+9\cos\left(  2t\right)  +\cos\left(  3t\right)  \right)
/15,$\\
$a_{2}=\left(  -5+3\cos\left(  2t\right)  +\cos\left(  3t\right)  \right)
/30,$\\
$a_{3}=2\left(  \left(  7+8\cos\left(  t\right)  \right)  \sin\left(
t/2\right)  ^{4}\right)  /15,$%
\end{tabular}
\]
and%
\[
\max_{t\in\mathbb{R}^{+}}\left(  \left\vert a_{j}\right\vert \right)  \left\{
\begin{tabular}
[c]{rr}%
$=1,$ & $j=1$ and $t=2\pi;$\\
$\approx3/10,$ & $j=2$ and $t\approx1/4;$\\
$\approx13/50,$ & $j=3$ and $t\approx91/50.$%
\end{tabular}
\right.
\]
This last equation concludes the proof of the theorem. Notice that $U_{TC}%
(\pi)$ is $2\times2$ block diagonal: the diagonal entries are $-13/15$; the
off-diagonal ones $2/15$. The two blocks are Grover matrices. When $t=\pi/2$,
all entries in the off-diagonal blocks are equal to $i/15$.%

\section{Conclusions}

We have shown that the 3-dimensional cube is the only periodic, connected
cubic graph with PST. The proof is based on known results of graph theory and
basic definitions of quantum dynamics on graphs. Our proof goes through the
relevant cases, which can be interpreted as a systematical exploration of
periodic quantum dynamics on cubic graphs. A necessary and sufficient
condition for PST would give a shorter, more elegant proof. However, any
method for the same task requires the different cases. There is a small plus
in the approach adopted here: we have written down explicitly the unitary
matrices that specify the dynamics. The matrices are potentially useful in
further work. The well-known mathematical scenario of state transfer together
with various facts observed in this paper suggest some reflections. These may
be worth a mention.

\subsection{Extremality}

The literature on interconnection networks for telecommunications, parallel
computers and distributed systems contain many optimization scenarios dealing
with order/degree/diameter \cite{fe}. The most famous one is perhaps the
\emph{degree/diameter problem }\cite{mil}: given natural numbers $D$ and
$\Delta$, find the largest possible number of vertices $n_{\Delta,D}$ in a
graph of maximum degree $\Delta$ and diameter $\leq D$. The graphs achieving
the upper bound are known as \emph{Moore graphs}. Interestingly, in the
setting of PST, the original problem was somehow the opposite one \cite{ch}.
In fact, the main point does not seek to maximize the number of vertices
without compromising reachability, but to perform long distance communication,
with a minimum of physical resources. The number of particles equals the order
of the graph and it is therefore a resource. The order/distance problems for
state transfer are then a new scenario, where graphs with extremal properties
can be defined via some parameter associated to a dynamical process, instead
of a topological condition. The landscape becomes even richer, if we consider
more than a single excitation, and lift the analysis to the graph powers
defined in \cite{godr}. These considerations remind that the problem of
identifying extremal graphs with respect to state transfer is still a mostly
unexplored area of research.

\subsection{A notion of persistency}

As a generalization of the rule inducing a continuous-time random walk, the
operator $U_{G}(t)=e^{-iA(G)t}$ has been subject of intense study (see
\cite{mo} and the references therein). In analogy with random walks, given the
importance of these objects for constructing distributions, a principal
direction has pointed uniform sampling. Additionally, still from the
quantitative side, it is useful to deal with problems related to single
vertices, or pairs of vertices, more than the entire graph. For example, an
instance of relevant parameters is a quantum version of the hitting time. We
would like to propose a notion that intends to quantify how the fidelity for
state transfer between two vertices is constant in an interval, including
small fluctuations. In the previous section, we have seen that half of the
diagonal entries of $U_{C_{6}+K_{2}}(t)$ have the form $a_{1}=\left(
2+\cos\left(  t\right)  +2\cos\left(  2t\right)  +\cos\left(  3t\right)
\right)  /6$. When $t\in\{\pi/2,3\pi/2\}$, $a_{1}=0$. Roughly between $\pi/2$
and $3\pi/2$ there is a plateau: the value of the function is about $1/3$; it
is exactly $1/3$, for $t=\pi$. Starting the process from some vertex $i$, if
at time $t=\pi$ we sample from the distribution, we are going to obtain $i$
with probability $1/9$. This probability is not spicked, but fairly stable
around $\pi$. Given a graph $G$, the $\epsilon$-\emph{persistency} of a pair
of vertices $i,j\in V(G)$ is the length of the longest interval $T\subset
\lbrack0,2\pi]$ such that there exists a value $k$ for which $k-\epsilon
<\left\vert [U_{G}(t)]_{i,j}\right\vert <k+\epsilon$, with $\epsilon\geq0$,
for every $t\in T$. By looking at all vertices, one could define the maximum
persistency or the average persistency, if extending the definition in the
obvious ways. A process with higher persistency requires less clock precision
for sampling. A first sight reason for giving attention to persistency could
arise from a connection with energy transfer problems \cite{ca}. Let us add an
additional remark. Decoherence has been shown to behave as a natural smoothing
mechanism on probability amplitudes \cite{ke}. Because of this fact, does
decoherence increase persistency? With a similar scope but a on a different
line, is the presence of decoherence compatible with PST at all?

\subsection{Discrete probability transfer}

The evolution governed by the Hamiltonian $H_{XY}(G)$ is driven by the
exponentially smaller $U_{G}(t)$, when taking into account a single excitation
only. While $A(G)$ faithfully represents $G$, the matrix $U_{G}(t) $ does not
preserve its topological structure. In other words, $[U_{G}(t)]_{i,j}\neq0$
does not imply $[A(G)]_{i,j}=1$. A discrete evolution on $G$ could be defined
(in some cases \cite{ss}) by a unitary $W_{G}$, with $[W_{G}]_{i,j}\neq0$ if
and only if $[A(G)]_{i,j}=1$. Such a process does not describe the transfer of
a single excitation, but it only allows to create a probability distribution
supported by $V(G)$. The process is discrete and it follows the iteration
$W_{G}W_{G}^{t-1}\longmapsto W_{G}^{t}$. If there is $t\in\mathbb{N}$ such
that $\left\vert \langle j|W_{G}^{t}|i\rangle\right\vert =1$ then we have
\emph{perfect probability transfer} from vertex $i$ to vertex $j$. Unless we
make use of some kind of \emph{lifting} \cite{amb} (\emph{e.g.}, the
introduction of extra degrees of freedom), this is the closest analogue to the
continuous object $U_{G}(t)$, even if this one does not always exist. For
example, we can construct on $K_{4}$ the unitary below:
\[
W_{K_{4}}=\left[
\begin{array}
[c]{cccc}%
0 & -1 & 1 & 1\\
1 & 0 & -1 & 1\\
1 & -1 & 0 & -1\\
1 & 1 & 1 & 0
\end{array}
\right]  .
\]
However, it is simple to observe that each power of $W_{K_{4}}$ has one of the
two zero-patterns%
\[
\begin{tabular}
[c]{lll}%
$\left[
\begin{array}
[c]{cccc}%
0 & \ast & \ast & \ast\\
\ast & 0 & \ast & \ast\\
\ast & \ast & 0 & \ast\\
\ast & \ast & \ast & 0
\end{array}
\right]  $ & or & $\left[
\begin{array}
[c]{cccc}%
\ast & 0 & \ast & \ast\\
0 & \ast & \ast & \ast\\
\ast & \ast & \ast & 0\\
\ast & \ast & 0 & \ast
\end{array}
\right]  ,$%
\end{tabular}
\
\]
where $\ast$ denotes a generic nonzero entry. This is sufficient to show that
$W_{K_{4}}$ does not give perfect probability transfer in $K_{4}$. In more
complicated situations, the zero-patterns of matrix powers are not immediately
available to imply a general statement. To verify that a graph $G $ does not
enjoy the property, we should study the spectra of unitary matrices with the
same zero-pattern of $W_{G}$. Since the problem involves both spectra,
zero-pattern, and an optimization procedure, its flavour reminds of the matrix
analysis questions approached in \cite{boy} or various parametrizations coming
from graph matrices \cite{hol}. In our context, the use of semidefinite
programming techniques does not seem immediately useful, because the matrices
are not stochastic, but in fact unitary.

\bigskip

\emph{Acknowledgments. }I\ am supported by a Newton International Fellowship.
I am grateful to Matthew Russell for finding an important error in a previous
version of the paper and an anonymous referee for valuable comments. This
paper is dedicated to Anthony Sudbery in the occasion of his retirement.


\section*{References}

\begin{thebibliography}{99}                                                                                               %


\bibitem {ah}O. Ahmadi, N. Alon, I. F. Blake, I. E. Shparlinski, Graphs with
Integral Spectrum, \emph{Linear Algebra Appl.}, \textbf{430}:1 (2009), pp. 547-552.

\bibitem {amb}A. Ambainis, Quantum Random Walks -- New Method for Designing
Quantum Algorithms, \emph{SOFSEM 2008: Theory and Practice of Computer
Science,} \emph{LNCS}, \textbf{4910} (2008).

\bibitem {an}R. J. Angeles-Canul, R. Norton, M. Opperman, C. Paribello, M.
Russell, C. Tamon, On quantum perfect state transfer in weighted join graphs,
\emph{Preprint 2009. }arXiv:0909.0431v1 [quant-ph]

\bibitem {godr}K. Audenaert, C. D. Godsil, G. F. Royle, T. Rudolph, Symmetric
squares of graphs, \emph{J. Comb. Theory, Ser. B} \textbf{97}(1): 74-90
(2007). arXiv:math/0507251v1 [math.CO]

\bibitem {ba}K. Bali\'{n}ska, D. M. Cvetkovi\'{c}, Z. Radosavljevi\'{c}, S.
Simi\'{c}, D. Stevanovi\'{c}, A survey on integral graphs, \emph{Univ.
Beograd, Publ. Elektrotehn. Fak., Ser. Mat. }\textbf{13} (2002), 42-65.

\bibitem {god1}A. Bernasconi, C. Godsil, S. Severini, Quantum networks on
cubelike graphs, \emph{Phys. Rev. A} \textbf{78}, 052320 (2008).
arXiv:0808.0510v1 [quant-ph]

\bibitem {blu}R. Bluhm, A. Kostelecky, J. Porter, B. Tudose, Revivals of
Quantum Wave Packets, COLBY 97-09, IUHET 372, September 1997, \qquad arXiv:quant-ph/9711061v1

\bibitem {boy}S. Boyd, P. Diaconis, J. Sun, and L. Xiao, Fastest mixing Markov
chain on a path, \emph{Amer. Math. Monthly}, \textbf{113} (2006), pp. 70--74.

\bibitem {bo}S. Bose, Quantum Communication through Spin Chain Dynamics: an
Introductory Overview, \emph{Contemporary Physics}, \textbf{Vol. 48} (1), pp.
13-30, 2007. arXiv:0802.1224v1 [cond-mat.other]

\bibitem {bos1}S. Bose, A. Casaccino, S. Mancini, S. Severini, Communication
in XYZ All-to-All Quantum Networks with a Missing Link, \emph{Int. J. Quantum
Info.}, \textbf{7}:3 (2009). arXiv:0808.0748v1 [quant-ph]

\bibitem {bc}F. C. Bussemaker and D. M. Cvetkovi\'{c}, There are exactly 13
connected, cubic, integral graphs, \emph{Univ. Beograd, Publ. Elektrotehn.
Fak., Ser. Mat. Fiz., Nos. }\textbf{544-576} (1976), 43-48.

\bibitem {ca}F. Caruso, A. W. Chin, A. Datta, S. F. Huelga, and M. B. Plenio,
Highly efficient energy excitation transfer in light-harvesting complexes: The
fundamental role of noise-assisted transport, \emph{J. Chem. Phys.},
\textbf{131} (10):105106, 2009. arXiv:0901.4454v2 [quant-ph]

\bibitem {ch}M. Christandl, N. Datta, A. Ekert and A. J. Landahl, Perfect
state transfer in quantum spin networks, \emph{Phys. Rev. Lett.} \textbf{92},
(2004), 187902. arXiv:quant-ph/0309131v2

\bibitem {con}M. Conder, P. Dobcs\'{a}nyi, Trivalent Symmetric Graphs Up to
\emph{768} Vertices, \emph{J. Combin. Math. Combin. Comput.} \textbf{40},
41-63, 2002.

\bibitem {cds}D. M. Cvetkovi\'{c}, M. Doob, H. Sachs, \emph{Spectra of graphs
-- Theory and application. }Deutscher Verlag der Wissenschaften -- Academic
Press, Berlin-New Tork, 1980.

\bibitem {fe}D. Ferrero, Introduction to interconnection network models,
\emph{Publ. Mat. Urug.}, 99/25 (1999).

\bibitem {god}C. Godsil, Periodic graphs, \emph{Preprint 2008}.
arxiv.org/abs/0806.2074 [quant-ph]

\bibitem {hol}H. van der Holst, L. Lov\'{a}sz, and A. Schrijver, The Colin de
Verdiere graph parameter, Graph theory and combinatorial biology
(Balatonlelle, 1996), \emph{Bolyai Soc. Math. Stud}, \textbf{7}, Janos Bolya
math. Soc. Budapest (1999), 29-85.

\bibitem {ke}V. Kendon, Decoherence in quantum walks - a review, \emph{Math.
Struct. in Comp. Sci} \textbf{17}(6) pp. 1169-1220 (2006). arXiv:quant-ph/0606016v3

\bibitem {mil}M. Miller, J. \v{S}ir\'{a}\v{n}, Moore graphs and beyond: A
survery of the degree/diameter problem, \emph{Elec. J. Comb.} (2005), \#DS14.

\bibitem {mo}M. Mosca, \emph{Quantum Algorithms}, Springer Encyclopedia of
Complexity and Systems Science (Springer, New York, 2009). arXiv:0808.0369v1 [quant-ph]

\bibitem {ore}O. Ore, \emph{Theory of graphs}, American Mathematical Society, 1962.

\bibitem {ru}M. Russell, Personal communication, January 2010.

\bibitem {s}A. J. Schwenk, Exactly thirteen connected cubic graphs have
integral spectra. \emph{Proc. Int. Graph Thory Conf.} \emph{at Kalamazoo}, May
1976, (Y. Alavi and D. Licks, Eds.) Springer-Verlag.

\bibitem {ss}S. Severini, On the digraph of a unitary matrix, \emph{SIMAX,
SIAM J. Matrix Anal. Appl.}, \textbf{25}, 1 (2003), pp. 295-300. arxiv:math.CO/0205187.

\bibitem {ka}W. Tadej, K. \.{Z}yczkowski, A concise guide to complex Hamadard
matrices, \emph{Open Syst. Inf. Dyn.} 13, 133-177 (2006). arXiv:quant-ph/0512154v2
\end{thebibliography}
\end{document}